\documentclass{jfm}

\usepackage{graphicx}
\usepackage{natbib}
\usepackage{dcolumn}
\usepackage{bm}
\usepackage{hyperref}
\usepackage{amsmath}
\usepackage{subfigure}
\usepackage{verbatim}

\usepackage{color}

\newcommand{\Snl}{S_{\text{nl}}}
\newcommand{\Snld}{S_{\text{nl,d}}}
\newcommand{\Sds}{S_{\text{ds}}}
\newcommand{\Sin}{S_{\text{in}}}

\makeatletter

\gdef\@underjournal{}

\def\footerflagdefns#1{}
\def\absfooterflag{}
\def\pagelimitfooter{}

\def\ps@titlepage{\leftskip\z@\let\@mkboth\@gobbletwo\vfuzz=5\p@
  \def\@oddhead{\vbox{\vspace*{-1pc}\hbox to \textwidth{\@j@urnal \hfil\llap{\thepage}}\par\vskip4pt}}
  \def\@evenhead{\@j@urnal \hfil\llap{\thepage}}%
  \def\@oddfoot{}%
  \def\@evenfoot{}%
  \def\sectionmark##1{}%
  \def\subsectionmark##1{}%
}

\def\ps@spheadings{\let\@mkboth\markboth
  \def\@oddhead{\hfil{\itshape\@shorttitle}\hfil\llap{\thepage}}%
  \def\@evenhead{\rlap{\thepage}\hfil\itshape\@shortauthor\hfil}%
  \def\@oddfoot{}%
  \def\@evenfoot{}%
  \def\sectionmark##1{\markboth{##1}{}}%
  \def\subsectionmark##1{\markright{##1}}%
}

\RequirePackage{etoolbox}
\patchcmd{\@maketitle}{(Received xx; revised xx; accepted xx)\hfill}{}{}{}

\makeatother

\title{A consistent generalised kinetic equation for damped ocean waves}

\author{Rhys Ransome\aff{1}, Alberto Alberello\aff{1,2}, Davide Proment\aff{1,3,4} \and Raphael Stuhlmeier\aff{5}}

\affiliation{\aff{1} School of Engineering, Mathematics and Physics, University of East Anglia, Norwich Research Park, Norwich, NR4 7TJ, United Kingdom
\aff{2} Department of Engineering \& Centre for Energy, Environment, and Sutainable Engineering, University of Huddersfield, Huddersfield, HD1 3DH, United Kingdom
\aff{3} Centre for Photonics and Quantum Science, University of East Anglia, Norwich Research Park, Norwich, NR4 7TJ, United Kingdom
\aff{4} ExtreMe Matter Institute EMMI, GSI Helmholtzzentrum fuer Schwerionenforschung, Planckstrasse 1, 64291 Darmstadt, Germany
\aff{5} Faculty of Civil \& Environmental Engineering, Technion -- Israel Institute of Technology, Technion City, Haifa, 3200003, Israel}

\begin{document}
\maketitle

\begin{abstract}
Contemporary spectral wave models, including those used in operational wave forecasts and in coupled earth system models, assume damping and non-linear wave energy exchange to be independent. Here, we critically examine this assumption. Starting from the equations of weakly non-linear waves, we {rigorously} derive a governing equation that consistently couples damping and non-linear interactions. This {consistent generalised kinetic equation} does not lead to an {appreciable} change in total wave--energy {dissipation}, but compared to the standard additive damping approach there is a noticeable difference in the distribution of energy among the wave frequencies. {This effect is particularly important for weak damping, and is a consequence of a memory effect retained in model.} Even stronger differences are observed when damping is allowed to depend on wave frequency, as is characteristic of many wave processes of growing interest such as the interaction with other components of the earth system, e.g. the cryosphere. Compared to the purely additive formulation, our new damped generalised kinetic equation has improved numerical stability characteristics.
 The formulation presented therefore provides a {possible} basis for future operational wave modelling. {While our results are for ocean waves, the derived generalised kinetic equation can be applied to other weakly nonlinear dispersive systems where external damping (or forcing) applies.}
\end{abstract}

\section{Introduction}
\label{sec:Introduction}

Ocean wave modelling has become one of the work-horses of modern marine science. Such modelling encapsulates the basic physics behind ocean surface waves, starting with their generation by the wind, through their nonlinear coupling, to their dissipation by white-capping or other processes, such as interaction with sea--ice or the sea--bottom. Today, wave models like WAVEWATCH III or SWAN are a key component of modern coupled Earth system models and an indispensable tool in weather forecasts, sediment transport studies, marine ecology, and a host of offshore operations \citep{Cavaleri2007,Cavaleri2012,Kousal2026}. Their products -- such as significant wave height or mean wave period -- are stochastic in nature, independent of individual wave phases, and valid over large spatial and temporal scales.

Such models are commonly referred to as spectral wave models, since they deal fundamentally with the averaged Fourier spectra of ocean waves. Indeed, the Fourier transform of the water surface can be directly related to the energy of the waves, already identified by \citet{Sverdrup1947} as the key variable for wave forecasts. With the advance of digital signal processing, \citet{pierson1952unified} and others began to reformulate the forecasting problem in the language of Fourier analysis. 

Fourier analysis is fundamentally a tool for linear systems, and it has been long understood that water waves behave in essentially nonlinear ways  \citep{Stokes1847}.  While significant mathematical advances in nonlinear waves were made in the second half of the 19th century, almost all of these dealt with steady, periodic wave motion. A first attempt to extend a higher-order theory of irregular waves -- with multiple Fourier harmonic components -- was made by \citet{Tick1959}, beginning an incredibly fruitful decade in nonlinear ocean wave research.

The discovery of the resonant energy exchange between quartets of deep-water waves by \citet{Phillips1960} laid the foundation for the derivation of the first kinetic equation for ocean surface waves by \citet{Hasselmann1962a} two years later. Inspired by developments in plasma physics and solid state theory (see \cite{Peierls1955} and \cite{Kadomtsev1965} for a review), and viewing the wave-field as weakly non-Gaussian -- in analogy with the idea of weak nonlinearity in the sense of Stokes --  Hasselmann derived an equation describing nonlinear energy transfer in a wave spectrum, thereby paving the way for modern, third-generation spectral wave modelling.

For quartets of waves in exact resonance, such that the algebraic sum of the four wavenumbers and frequencies vanishes, Hasselmann's kinetic equation describes the evolution of the wave energy spectrum on a slow time-scale $T/\epsilon_0^4,$ for $T$ a typical wave period and $\epsilon_0$ a typical steepness of the sea-state. While this represented a major theoretical advance, the practical significance of these results was not immediately evident.  

Nonlinear energy exchange is one mechanism that influences the growth and propagation of ocean waves, but by no means the most important. In fact, first-generation wave models were built entirely around the idea that (spectral) wave energy $C$ is transported at the deep-water group velocity $\mathbf{v},$ 
\[ \frac{\partial C}{\partial t} + \mathbf{v} \cdot \nabla C = A + B \cdot C + \Sds \]
with an external turbulent pressure forcing $A$ \citep{phillips1958equilibrium} and a linear feedback mechanism $B\cdot C$ \citep{Miles1957} simulating wave-growth by wind, alongside an on/off limiter $\Sds$ to prevent the spectrum from exceeding a specified saturation level. Together the right-hand side represented what are still called input and dissipation source terms, and denoted $\Sin$ and $\Sds$. 

Observed deficits in the ability of these simple models to describe the measured growth of wind waves, together with doubts about their representation of the underlying physics, led to a re-evaluation of the spectral energy balance in the 1970s \citep{SWAMP1985}. Nevertheless, computational implementation of the nonlinear source term   $\Snl$ continued to pose challenges. 
The fundamental problem is that the nonlinear interaction integral must be taken over a three-dimensional wavenumber space leading to a five-dimensional wave evolution model, and finding the resonant manifold is computationally demanding.  Serious efforts to overcome this hurdle bore fruit in the 1980s, particularly with the work of \cite{HasselmannHasselmann1985,Hasselmann1985} who developed the Direct Interaction Approximation, culminating in the implementation of the kinetic equation in a so-called third-generation wave model by \cite{WAMDI_1988} (more details of which can be found in \citet{komen1996dynamics}). A number of important numerical improvements were made in subsequent decades, such as the Webb-Resio-Tracy (WRT) method \citep{Webb1978,VanVledder2006}, which today are implemented in state-of-the-art wave forecasting models like WAVEWATCH III \citep{WW3DG2016}.

As Hasselmann's kinetic equation went from theoretical innovation to a well-integrated part of operational models, research increasingly began to probe the limits of its validity. The advent of efficient, large-scale direct numerical simulations \citep{tanaka2001verification} allowed researchers to examine some of the underlying assumptions, such as the primacy of exact resonant interactions. Indeed, in Hasselmann's equation, off--resonant interactions should not contribute to the spectral evolution, a fact related to the long-time integration involved in the derivation. 

Work by \citet{Annenkov2001a} and \citet{tanaka2001verification} among others critically assessed these assumptions, and a first generalisation of the kinetic equation to account for near-resonant interactions operating on faster time-scales was made by \citet{Janssen2003}. Subsequent work by Annenkov \& Shrira using direct numerical simulations of the primitive Zakharov equation highlighted the fact that near-resonant contributions are every bit as important as exact resonances in determining spectral evolution, and also proposed a generalisation of Janssen's equation \citep{Annenkov2006}. This was further extended in work of \citet{Gramstad2013} to include Stokes frequency corrections. 

These generalised kinetic equations (GKE) dispense with the long-time integration of Hasselmann's formulation, and instead include a memory integral over the nonlinear interaction terms. Consequently, they describe evolution on the fast $T/\epsilon_0^2$ and slow $T/\epsilon_0^4$ (kinetic) time-scales, and have been used in recent years to model transient behaviour \citep{Annenkov2016} associated with wind forcing, compared with Hasselmann's equation \citep{Annenkov2018} and tested in wave models \citep{Gramstad2016,Liu_Gramstad_Babanin_2021}. One drawback of these generalised models is that they are computationally costlier than the usual kinetic wave models, as the resonant condition over the wave angular frequency is not enforced, leading to an integral in four-dimensional space, while the memory term also increases the complexity of the time integration.

When the nonlinear contributions are implemented in wave models, the combined ``source term" is usually written 
\begin{equation} \label{eq:source-term-splitting} S = \Sin + \Sds + \Snl,\end{equation}
including representations of wind input $\Sin$ and dissipative processes $\Sds$ such as whitecapping, as well as four-wave resonant interactions. Numerous semi-empirical formulations of the source terms are available nowadays \citep{Ardhuin2010,Liu2019}, noting that for many physical processes their strength is function of frequency and spectral shape.  The nonlinear interaction term $\Snl$ is not strictly a source of energy, but rather serves to redistribute it within the spectrum. The source terms are thus assumed to be mutually non-interacting, and the nonlinear term $\Snl$ to be independent of the linear terms $\Sds$ and $\Sin$ based on an argument of separation of scales between the distinct physical processes \citep{hasselmann1974spectral}.

Recently, the validity of this assumption has been reassessed by \citet{maestrini2025wave}, who derived Hasselmann's kinetic equation directly from a deterministic primitive equation with damping. In the limit of zero nonlinearity, when damping evolves on a slow kinetic time scale, the additive formulation \eqref{eq:source-term-splitting} is recovered. On the faster damping time scale, nonlinearity $S_{\text{nl}}$ is modified, and the formulation is no longer additive.  

When assumptions about spectral evolution are relaxed to allow near resonant contributions, the resulting generalisations of Hasselmann's kinetic equation account for spectral evolution on both slow and fast time-scales. This raises the question whether the simple summing of source terms is appropriate in this context. 
To study the interplay between the damping, near resonant and kinetic time scales, we derive a novel generalised kinetic equation in the presence of damping directly from the primitive equations, and explore its behaviour via numerical simulation.

In Section \ref{sec:Mathematical Formulation} we present the basic formulation of the problem, covering the derivation of the damped generalised kinetic equation and its numerical implementation. In Section \ref{sec:uniform-damping-results} we present results for a uniform, frequency-independent damping. Section \ref{sec:frequency-dependent-damping} considers the more realistic case where damping depends on wave frequency, and shows simulations of spectral evolution for non-uniform damping. Section \ref{sec:Discussion and conclusions} provides a discussion of our results and concluding remarks.

\section{Mathematical Formulation }
\label{sec:Mathematical Formulation}

\subsection{Nonlinear wave-wave interactions in deep water}

The kinetic theory of surface waves is derived from a perturbation expansion of the inviscid, irrotational water wave problem. In deep water, surface gravity waves are dispersive, and mutual resonant interaction appears only at cubic nonlinearity. Therefore, the preferred starting point \citep{Janssen2003,Mei2005} for deriving the kinetic equation is the cubic Zakharov equation, 
\begin{equation} \label{eq:ZE-continuous}
    i \frac{db(k)}{dt}=\omega(k) b(k) + \iiint_{-\infty}^{\infty} T(k,k_1,k_2,k_3)b^*(k_1)b(k_2)b(k_3)\delta(k+k_1-k_2-k_3) dk_1 dk_2 dk_3.
\end{equation}
derived from the Hamiltonian formulation of the water wave problem by \citet{Zakharov1968}.

The complex amplitude $b(k,t)$ is related to the spatial Fourier transform of the free surface $\zeta(x,t)$ by 
\begin{equation} \label{eq:h-zeta-from-b-k}
    \hat{\zeta}(k,t) = \sqrt{\frac{\omega}{2g}} \left( b(k) + b^*(-k) \right)
\end{equation}
to lowest order, where $\omega^2 = g|k|$ is the radian frequency in deep water, $g$ is the acceleration of gravity, $k$ the (scalar or vector) wavenumber, $*$ denotes a complex conjugate, and the function $T$ is the interaction kernel (see \citet{Krasitskii1994}). This remarkably simple equation describes the deterministic evolution of deep-water waves to third-order, generalising Stokes' perturbation expansion for steady waves (see \cite{Stuhlmeier2024}) as well as prominent model equations like the nonlinear Schr\"odinger equation (NLS) \citep{Mei2005} or Dysthe equation \citep{Stiassnie1984c}.

A deterministic model is impractical for large-scale wave forecasts, since this would require information about individual phases which is typically not available (although on the small-scale, phase-resolved forecasting using \eqref{eq:ZE-continuous} is feasible \citep{Stuhlmeier2021}). Moreover, the stochastisation of surface waves \citep{Annenkov2001} imposes dynamic constraints in addition to the geometric limitations inherent to deterministic forecasts. For this reason, we seek to model the evolution of the variance density spectrum of the sea surface $\zeta$, defined as the Fourier transform of the two-point correlation 
\[ S(k) = \mathfrak{F}_{\xi \mapsto k} \left\lbrace \langle \zeta(x+\xi) \zeta(x)  \rangle \right\rbrace, \]
where the brackets denote an ensemble average, and $\mathfrak{F}_{\xi \mapsto k}$ denotes the $\xi$ to $k$ Fourier transform. For this to be a sensible definition, we must assume that the free surface is spatially (weakly) homogeneous, so that the autocorrelation is independent of spatial location $x$, and depends only on separation distance $\xi.$ 

With this definition of the variance density spectrum, it is possible to define a number of averaged metrics that characterise the wave field \citep{Holthuijsen2007}. Most prominently, the zeroth spectral moment
\begin{equation} m_0 = \int_0^\infty S(k) \text{d} k, \end{equation}
which is related to the (spectral) significant wave-height $H_{m_0}$ by $H_{m_0} \approx 4 \sqrt{m_0}.$ Here and below, for simplicity, we will restrict ourselves to unidirectional wavenumber spectra, although our derivations remain valid for directional spectra. 

Introducing a discretisation in Fourier space, with wavenumber spacing denoted $dk,$ we can then establish the relationship between the complex amplitudes $b_i=b(k_i)$ and the variance density spectrum $S(k)$ as 
\begin{equation} \label{eq:C_i-to-S_i relation}
\langle b_i b_i^* \rangle = \frac{g}{\omega_i} S(k_i) dk.
\end{equation}
Henceforth we will use sub- and superscripts to denote wavenumber indices where convenient, abbreviating terms such as $\omega(k)$ by $\omega_k,$ $T(k,k_1,k_2,k_3)$ as $T_{0123}$ or $\delta(k_n+k_p-k_q-k_r)$ as $\delta_{np}^{qr}.$

A discrete formulation is essential for numerical simulations, and implies a finite physical domain $D = [0,L],$ which in turn sets the discretisation of the Fourier space as $\Delta k = 2\pi/L$. We rewrite the Zakharov equation \eqref{eq:ZE-continuous} in discrete form as
\begin{equation} \label{eq:ZE-discrete-damping}
\frac{d b_n}{dt} = -i \omega_n b_n - \mu_n  b_n - i \sum_{p,q,r} T_{npqr} b^*_p b_q b_r \delta_{np}^{qr},
\end{equation}
where we have included an external forcing or damping term denoted by $\mu_n b_n$ where the sign of  $\mu_n$ dictates whether forcing ($\mu_n<0$) or damping ($\mu_n>0$) is at play. The magnitude of $\mu_n$ also dictates the time-scale on which the forcing/damping acts in comparison to the linear and nonlinear terms \citep{maestrini2025wave}, as shall be explored in more detail. While both signs of $\mu_n$ are mathematically equivalent, we will focus on cases with damping in the results presented below.

\subsection{Derivation of a damped generalised kinetic equation}

Equation \eqref{eq:ZE-discrete-damping} is the starting point for our derivation of a generalised kinetic equation, which proceeds along the lines of work by \citet{Gramstad2013}. Under the assumption of spatial homogeneity of the wave field, we compute the evolution of the wave action spectrum $C_n = \langle  |b_n|^2 \rangle$, using linearity of the expected value to write 
\begin{equation} \label{eq:dCn/dt-intermediate-1}
\frac{dC_n}{dt }  =   - 2\mu_n C_n +2\text{ Im} \sum_{p,q,r} T_{npqr} \langle b_n^* b_p^* b_q b_r \rangle \delta_{np}^{qr}.
\end{equation}

Equation \eqref{eq:dCn/dt-intermediate-1} presents the usual challenge encountered in weak wave turbulence {due to nonlinearity} \citep{Nazarenko2011}: the evolution of a second-order average depends on a fourth-order average, the start of a hierarchy which continues indefinitely. Since linear wave fields are described by Gaussian processes \citep{Trulsen2004}, and our wave theory is weakly nonlinear up to third order -- in the sense that the perturbation expansion presents a small correction to the linear theory -- it is appropriate to assume that departures from Gaussian statistics are likewise small. 

In this vein, we may write the fourth and sixth order correlators, introducing the fourth order joint cumulant $\kappa_{npqr}$ and the sixth-order cumulant $\kappa_{npqruv}$, 
\begin{align}   \label{eq:4th-order-correlator-cumulant-relation}
\langle b_n^* b_p^* b_q b_r \rangle & = C_n C_p ( \delta_n ^q \delta_p^r + \delta_n ^r \delta_p ^q) + \kappa_{npqr}, \\
\left\langle b_n^{*} b_p^{*} b_q^{*} b_r b_u b_v \right\rangle
& = C_n C_p C_q
\left(
\delta_n^{r}\delta_p^{u}\delta_q^{v}
+\delta_n^{r}\delta_p^{v}\delta_q^{u}
+\delta_n^{u}\delta_p^{r}\delta_q^{v}
+\delta_n^{u}\delta_p^{v}\delta_q^{r}
+\delta_n^{v}\delta_p^{r}\delta_q^{u}
+\delta_n^{v}\delta_p^{u}\delta_q^{r}
\right) \nonumber
\\
&+
C_n
\left(
\kappa_{pquv}\,\delta_n^{r}\delta_{pq}^{uv}
+\kappa_{pqrv}\,\delta_n^{u}\delta_{pq}^{rv}
+\kappa_{pqru}\,\delta_n^{v}\delta_{pq}^{ru}
\right) \nonumber
\\
&+
C_p
\left(
\kappa_{nquv}\,\delta_p^{r}\delta_{nq}^{uv}
+\kappa_{nqrv}\,\delta_p^{u}\delta_{nq}^{rv}
+\kappa_{nqru}\,\delta_p^{v}\delta_{nq}^{ru}
\right) \nonumber
\\
&+
C_q
\left(
\kappa_{npuv}\,\delta_q^{r}\delta_{np}^{uv}
+\kappa_{nprv}\,\delta_q^{u}\delta_{np}^{rv}
+\kappa_{npru}\,\delta_q^{v}\delta_{np}^{ru}
\right)
+\kappa_{npqruv},
\end{align}
and posit $\kappa_{npqr} = o(\epsilon^4), \kappa_{npqruv} = o(\epsilon^6)$. Note that wave-field homogeneity is enforced by the $\delta$-functions appearing in this expansion.

Defining the frequency detuning
\begin{equation} 
\Delta_{np}^{qr} = \omega_n + \omega_p - \omega_q - \omega_r,
\end{equation}
which vanishes in exact resonance, and the related quantity
\begin{equation} 
\Delta \mu_{npqr} = \mu_n + \mu_p + \mu_q + \mu_r.
\end{equation}
we use \eqref{eq:ZE-discrete-damping} to obtain the evolution of the fourth-order correlator:
\begin{align} 
\frac{d}{dt}\langle b_n^* b_p^* b_q b_r \rangle  
&= \nonumber
\left( i \Delta_{np}^{qr} -  \Delta \mu_{npqr} \right)
\langle b_n^* b_p^* b_q b_r \rangle \\ \nonumber
&+ i \sum_{u,v,w} T_{nuvw}
\langle b_p^* b_v^* b_w^* b_q b_r b_u \rangle
\delta_{nu}^{vw}
+ i \sum_{u,v,w} T_{puvw}
\langle b_n^* b_v^* b_w^* b_q b_r b_u \rangle \nonumber
\delta_{pu}^{vw}\\
&- i \sum_{u,v,w} T_{quvw}
\langle b_n^* b_p^* b_u^* b_r b_v b_w \rangle
\delta_{qu}^{vw}
- i \sum_{u,v,w} T_{ruvw}
\langle b_n^* b_p^* b_u^* b_q b_v b_w \rangle
\delta_{ru}^{vw}.
\end{align}
By differentiating in time the relation \eqref{eq:4th-order-correlator-cumulant-relation} between the fourth order cumulant and the four point correlator, one obtains
\begin{equation} 
\frac{d}{dt}\langle b_n^* b_p^* b_q b_r \rangle = \left( \frac{dC_n}{dt} C_p + C_n \frac{d C_p}{dt} \right) ( \delta_n ^q \delta_p^r + \delta_n ^r \delta_p ^q) + \frac{d\kappa_{npqr}}{dt}.
\end{equation}

Gathering the preceding equations, we may therefore write the time derivative of the fourth order cumulant as
\begin{equation} \label{eq:d-kappa-fourth-order}
\frac{d \kappa_{npqr}}{dt} = ( i \Delta_{np}^{qr} -  \Delta \mu_{npqr}) \kappa_{npqr} + i \alpha_{npqr} + i\beta_{npqr} \kappa_{npqr},
\end{equation}
where 
\begin{align} 
\alpha_{npqr}&= 2T_{npqr}\delta_{np}^{qr} [ C_q C_r (C_n + C_p) - C_n C_p (C_q +C_r] = 2T_{npqr}\delta_{np}^{qr}f_{npqr}, \\
\beta_{npqr}&= 2 \sum_u C_u ( T_{nunu } + T_{pupu} - T_{ququ} - T_{ruru}),
\end{align}
and
\begin{equation}
     f_{npqr}= C_q C_r (C_n + C_p) - C_n C_p (C_q +C_r) = C_nC_pC_qC_r \left( \frac{1}{C_n}+ \frac{1}{C_p} - \frac{1}{C_q}-\frac{1}{C_r}\right).
\end{equation}

Rearranging equation \eqref{eq:d-kappa-fourth-order}
\begin{equation} \label{eq:cumulant-evolution}
\frac{d \kappa_{npqr}}{dt} - \left[ i(\Delta_{np}^{qr} + \beta_{npqr}) - \Delta \mu_{npqr})\right] \kappa_{npqr} = i \alpha_{npqr}
\end{equation}
and defining
\begin{equation} 
\theta(t) = \Delta_{np}^{qr} t + \int_0^t \beta_{npqr}(\lambda) d \lambda
\end{equation}
as standard, we assume a cold start, i.e, the initial value of the cumulant $\kappa_{npqr}(0) =0$,  allowing us to integrate in time:
\begin{equation} \label{eq:cumulant-evolution-integrated}
\kappa_{npqr} e^{-i \theta(t)+  \Delta \mu_{npqr }t} = i \int_0 ^t \alpha_{npqr}(t') e^{-i \theta(t')+ \Delta\mu_{npqr }t'}\,  d t'.
\end{equation}
Finally, substituting this result together with \eqref{eq:4th-order-correlator-cumulant-relation} into \eqref{eq:dCn/dt-intermediate-1} 
yields
\begin{equation} \label{eq:dGKE}
\frac{dC_n}{dt } =  \underbrace{- 2 \mu_n C_n \vphantom{\sum_p}}_{\Sds } +\underbrace{4\text{ Re} \sum_{p,q,r} T^2_{npqr}  \delta_{np}^{qr} \int_0 ^t f_{npqr}(t') e^{i (\theta(t) - \theta(t'))} e^{- \Delta\mu_{npqr }(t-t')}\,  d t'}_{S_\text{nl,d}}.
\end{equation}

This generalised kinetic equation, with damping implemented in a consistent manner, \eqref{eq:dGKE} is our main theoretical result. One of its main features, which differentiates it from Hasselmann's formulation, is the memory integral appearing on the right-hand side. 
With our rigorous derivation, it is clear that the damping appears in two places. First in the term $-2\mu_n C_n$ which is linear in the spectrum, and agrees with the way damping is usually modelled as an additive source term ($S_{\text{ds}}$). Second, in the nonlinear term ($S_{\text{nl,d}}$)  as an exponential term involving $\Delta \mu_{npqr}$ in the memory integral, which could not be anticipated by the \textit{a posteriori} addition of damping to the conservative equation. We refer to this new form of the nonlinear source term in the presence of damping as $S_{\text{nl,d}}.$

One further difference between Hasselmann's kinetic equation and the generalised formulation is the fact that the latter allows for a (weak) build-up of correlations among the phases due to the memory term. This stems from retaining the cumulant evolution with time \eqref{eq:cumulant-evolution}, rather than treating this integral as stationary. Consequently, if the value of the cumulant $\kappa_{npqr}$ is set to zero initially, correlations among the phases accumulate with time according to \eqref{eq:cumulant-evolution-integrated}, representing departures from Gaussian statistics. The real part of the cumulant $\kappa_{npqr}$ can be linked to the dynamic kurtosis of the wave-field \citep{Annenkov2016,ONORATO20163173}. By contrast, Hasselmann's kinetic equation expresses the cumulant $\kappa_{npqr}$ explicitly in terms of the instantaneous spectral density $f_{npqr}$, thereby precluding the accumulation of phase correlations. 

Prior work on the generalised kinetic equation \citep{Gramstad2013,Gramstad2016,Liu_Gramstad_Babanin_2021} has argued that realistic wave evolution is accompanied by periodic de-correlation of phases due to processes such as wave breaking. While such a process cannot be analytically modelled within equation \eqref{eq:dGKE} itself, it is possible to implement it numerically by periodic re-setting of the fourth-order cumulants to zero, giving rise to fast adjustment in the GKE.

\subsection{Numerical implementation and setup}
\label{subsec:numerical-implementation}

Our numerical implementation of \eqref{eq:dGKE} follows that initially proposed by \citet{Gramstad2013}, extended here to account for dissipation. The interaction is discretized over a finite, 1D wavenumber grid, evolving the wave action density $C_n(t)$ for each discrete mode $k_n$. 
In practice we employ $N = 1024 $ wave modes for all calculations, with $k_n = n \Delta k$ ranging between 0 and $4 k_{p,0}$, for $k_{p,0}$ the peak wavenumber of our initial spectrum. To compute the nonlinear energy transfer, the scheme relies on a pre-computed interaction kernel $T_{npqr} $. The identification of interacting quartets can be made more efficient by using symmetries in the interaction ($n+p=q+r$), thereby reducing the number of redundant calculations and lowering memory usage.

Denoting the linear dissipation and nonlinear interaction terms by  $\mathcal{L}_n(\boldsymbol{C}) \text{ and } \mathcal{N}_n(\boldsymbol{C})$, and their respective sub-problems: $\dot{C}_n = \mathcal{L}_n(\boldsymbol{C})$ and $\dot{C}_n = \mathcal{N}_n(\boldsymbol{C})$, we maintain numerical stability and integration accuracy in the presence of  non-conservative damping terms using an operator splitting technique \citep{strang1968construction}. Rather than integrating the nonlinear transfer and the linear dissipation simultaneously using an explicit Euler method, the splitting isolates the linear terms into exact exponential half-steps. During each time step $\Delta t$, the wave action density is first advanced via a half-step of the exact analytical solution to the linear sub-problem (i.e. the damping $S_{\text{ds}}$), with the nonlinear term subsequently computed using the explicit scheme proposed by \citet{Gramstad2013} to solve the nonlinear sub-problem, $C_n \rightarrow C_n + \Delta C_n ^{{\text{NL}}} $, followed with a final exponential half-step application;

\begin{equation} \label{eq:strang-outline}
C_n ^* = C^s_n \exp(-\mu_n \Delta t) ,\, \, C_n^{**} = C_n^* +  \Delta C_n ^{{\text{NL}}} (  \boldsymbol{C}^*), \, \, C_n ^{s+1} = C^{**}_n \exp(-\mu_n \Delta t)
\end{equation}
where $C_n ^s = C_n(s\Delta t )$, and  $ \Delta C_n ^{{\text{NL}}}$ in our rigorous derivation differs by only needing to include the factor $e^{- \Delta\mu_{npqr }(t-t')}$ in the integrand compared to the additive formulation. This separation of the physical operators prevents the rapid exponential damping functions from degrading the stability constraints of the explicit nonlinear solver.

In order to better understand the effects of the newly derived $\Snld$ term in \eqref{eq:dGKE}, we compare our numerical solutions with those of the related damped GKE with purely additive damping term
\begin{equation} \label{eq:additive-GKE}
\frac{dC_n}{dt } =  \underbrace{- 2\mu_n C_n \vphantom{\sum_p}}_{S_{\text{ds}}} +\underbrace{4\text{ Re} \sum_{p,q,r} T^2_{npqr}  \delta_{np}^{qr} \int_0 ^t f_{npqr}(t') e^{i (\theta(t) - \theta(t'))} \,  d t'}_{S_{\text{nl}}}, 
\end{equation}
which is analogous to what is usually undertaken for Hasselmann's kinetic equation. To differentiate this from the rigorously derived formulation, we denote the nonlinear term in the absence of damping, which is identical to that obtained by \citet{Gramstad2013}, by $S_{\text{nl}}.$

As a reference case we also consider only linear damping in the absence of any interaction
\begin{equation} \label{eq:linear-decay}
    \frac{dC_n}{dt } =  \underbrace{- 2\mu_n C_n}_{S_{\text{ds}}},
\end{equation}
which will allow for a simple, visual comparison of the effects of damping and nonlinear interaction on the spectral shape. We will refer to this dissipation source term as $S_{\text{ds}}.$

Initial conditions for our numerical simulations will be in the form of a JONSWAP wavenumber spectrum 
\begin{equation}
    \label{eq:JONSWAP}
    S(k) = \frac{\alpha}{2 k^3} \exp\left[-\frac{5}{4} \left(\frac{k}{k_{p,0}}\right)^{-2}\right] \gamma^{\exp\left[ - \left( \sqrt{\frac{k}{k_{p,0}}}-1 \right)/2\sigma^2 \right]}.
\end{equation}
The spectrum is characterised by an energy scale parameter $\alpha,$ the peak wavenumber $k_{p,0}$, and the shape parameters $\gamma$ and $\sigma,$ where $\gamma$ represents a peak-enhancement parameter which modifies the Pierson-Moskowitz spectrum (in which $\gamma=1$) which describes fully developed seas. In all numerical simulations, we set $\alpha=0.0176$ and choose a peaked spectrum with $\gamma=6$ together with an initial $k_{p,0}=0.02 \text{ m}^{-1},$ such that the initial spectral wave steepness 
\[ \epsilon_0 = k_{p,0} \frac{H_{m_0}}{2}\]
takes the value $\epsilon_0 \approx 0.12$ (see \eqref{eq:C_i-to-S_i relation}), which is equivalent to the cases studied by \citet{Gramstad2013}. With these parameters, representative of energetic wind seas, the (initial) peak wavelength is $314 \text{ m}$ and the peak period is $T_{p,0} = 14.19 \,\text{s}.$ 

\subsubsection{Phase-mixing}

Because the build-up of correlations in the generalised kinetic equation may be periodically reset by physical processes in the sea (such as whitecapping), previous studies have employed numerical phase mixing in comparing GKE simulations to Hasselmann's kinetic equation \citep{Liu_Gramstad_Babanin_2021}. The periodic re-setting of the fourth-order cumulants also has a stabilising effect on the numerical solution. Therefore, we seek to compare simulations of the (additive and rigorous) damped GKE with and without phase-mixing. While a range of choices of phase-mixing intervals are possible, we elect to re-initialise the phases every 50 peak periods to facilitate comparison with the work of \citet{Gramstad2013}.

\subsubsection{Choice of dissipation strength}

To study the interplay between dissipation and nonlinear wave interactions for different damping rates, we vary the damping magnitude in our simulations. We will concentrate on three representative damping rates which span the rate of evolution of Hasselmann's kinetic equation $ \mathcal{O}(\epsilon_0^4 \omega_{p,0})$ to the rate of near-resonant dynamic interactions $ \mathcal{O}(\epsilon_0^2 \omega_{p,0})$ \citep{Janssen2003,Annenkov2006}. 

These damping rates are established by an action-weighted average over the initial spectrum, 
\begin{equation} \label{eq:forcing-average}
\langle \mu_n \rangle _C = \frac{\sum_n \mu_n C_n(0)}{\sum_n  C_n(0)},
\end{equation}
which allows for a sensible definition of average damping even when the dissipation varies with frequency. In the case of constant dissipation $\mu_n = \mu $, so that $\langle \mu_n \rangle _C = \mu$. The mean damping rates which will be used for our core calculations are  $\langle \mu_n \rangle _C  = 1.6 \epsilon_0^4 \omega_{p,0}, 1.1 \epsilon_0^3 \omega_{p,0}, 0.79 \epsilon_0 ^2\omega_{p,0}$.

\subsubsection{Quantifying changes in spectral shape}

In order to quantify the difference between the spectral shape in the different formulations, we compute the Hellinger distance $d_H$ \citep{hellinger1909neue} between the two normalised energy spectra $\hat{S}_i(k) , \hat{S}_j(k) $, defined as
\begin{equation} \label{eq:Hellinger-def}
d_H( S_i , S_j ) = \left[ \frac{1}{2} \int_0^{\infty}  \left( \sqrt{\hat{S}_i(k)}-\sqrt{\hat{S}_j(k)}\right)^2  dk\right]^{1/2}, \, \, \, \hat{S}(k) =  \frac{S(k)}{m_0}.
\end{equation}
Normalising the energy ensures that $d_H$ is a direct measure of spectral reshaping only, taking values $d_H \in [0,1]$, where $d_H=0$ and $d_H=1$ denote identical spectral shapes and no overlap in spectral shapes, respectively. 

\section{Results for uniform damping}
\label{sec:uniform-damping-results}

\subsection{Spectral evolution}
\label{ssec:Spectral evolution UD}
 
We will first study the simplified case of uniform linear damping $\mu_n = \mu$, where all frequencies in the spectrum are damped at equal rates.  While this is clearly not representative of real seas, it has the advantage that the equation with damping-only \eqref{eq:linear-decay} preserves the spectral shape.  Changes in spectral shape with time can therefore be attributed entirely to the nonlinear interaction, either in the usual, purely additive form $\Snl$ or the rigorously derived form $\Snld$ which is implemented in \eqref{eq:dGKE}.

Each initial configuration is numerically simulated using the rigorous ($\Sds+\Snld$) and purely-additive ($\Sds+\Snl$) formulations both with and without phase-mixing. A reference evolution is also computed with linear damping and no nonlinear interactions ($\Sds$ only), by solving equation \eqref{eq:linear-decay} analytically.

In comparing the time-evolution of spectra with different damping rates, these spectra should not be compared at a fixed time, but rather at a fixed energy. To this end, we adopt the damping-dependent integration time $t^*,$ selected so that the total spectral energy has reduced to 10 \% of its initial value i.e., 
\begin{equation} \label{eq:tStar-def}
m_0(t^*) = m_{0}(0)/10 \, .
\end{equation}
For the three cases considered $t^* = 585T_{p,0}, 99T_{p,0},$ and $ 16.8T_{p,0}$, respectively. The integration time--step is chosen as $\Delta t = t^*/4000.$

\begin{figure}
    \centering
    \includegraphics[width=\textwidth]{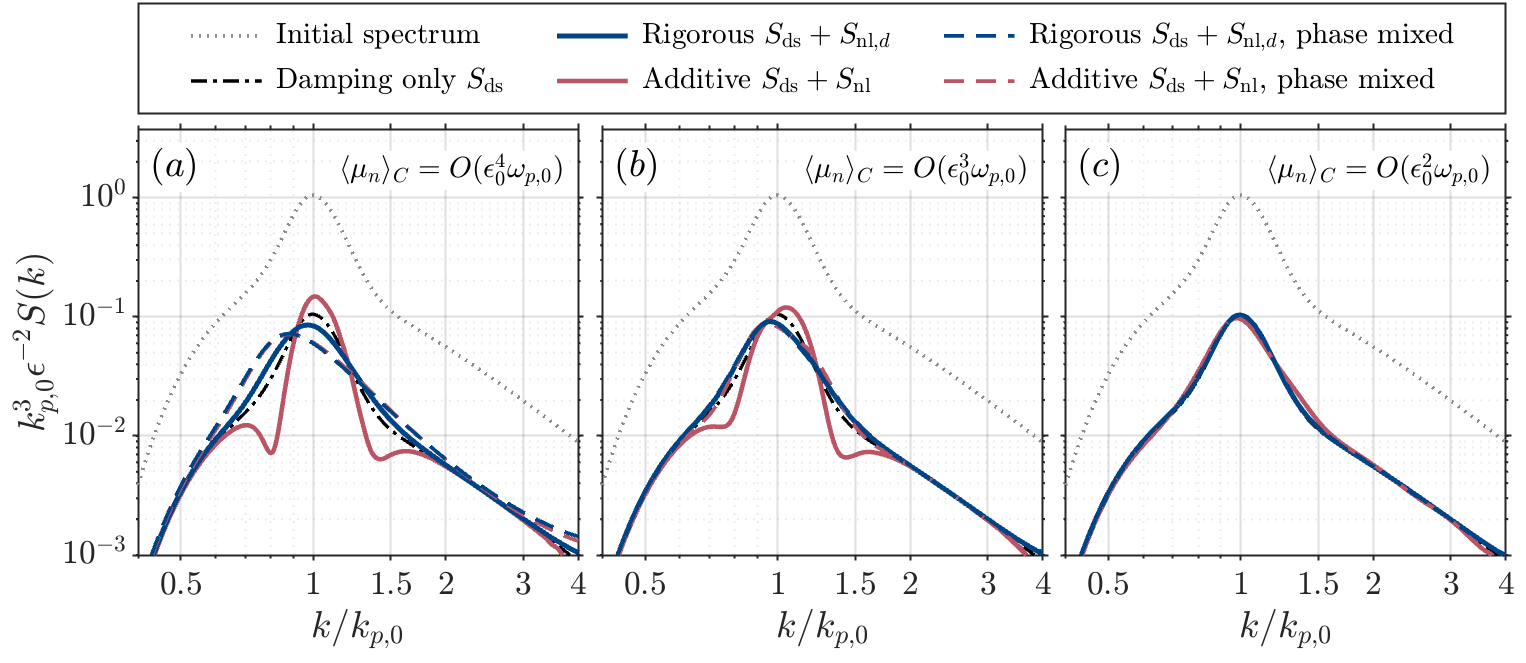}
    \caption{Dimensionless wavenumber spectra at $t^*$ for damping only, rigorous and additive with and without phase mixing shown for three constant damping rates in panels:  (a) $ \langle\mu_n \rangle_C  =  1.6 \epsilon_0^4 \omega_{p,0}$; (b)$ \langle\mu_n \rangle_C  =1.1 \epsilon_0^3\omega_{p,0}$; (c) $ \langle\mu_n \rangle_C  = 0.79 \epsilon_0 ^2\omega_{p,0}$.}
    \label{fig:spectra-equal-energy}
\end{figure}

Figure \ref{fig:spectra-equal-energy} shows the dimensionless linearly damped wavenumber spectrum at $t^*$ 
together with simulations based on the rigorous and purely-additive nonlinear terms alongside their phase-mixed counterparts
for the three uniform damping rates.
Since the damping is uniform, the damping-only solution to the linear equation \eqref{eq:linear-decay} is simply a scaled down initial spectrum (dash-dotted black curve), with the peak of $k_{p,0}^3 \epsilon_0^{-2}S$ falling from $1.05$ to $0.11$ at $k/k_{p,0} = 1$. 

The weakest damping $\mathcal{O}(\epsilon_0^4 \omega_{p,0})$, shown in Figure \ref{fig:spectra-equal-energy}(a), shows large differences between the rigorous (in blue) and purely-additive (in red) simulations when phase mixing is not implemented. 
Heuristically, the time-scale of the damping is slow enough to allow for significant nonlinear interaction (a point which is discussed more fully in Sec.\ \ref{ssec:Time scales}). The additive $\Snl$ yields a higher, sharper spectral peak compared to the damping-only case, with the appearance of local minima on either side of it. In contrast, the rigorous $\Snld$ exhibits a broader spectrum with lower peak and no local minima.

As the damping rate is increased to $\mathcal{O}(\epsilon_0^3 \omega_{p,0})$ so the spectral reshaping is reduced. Nevertheless, the same qualitative differences seen for weaker damping are visible in Figure \ref{fig:spectra-equal-energy}(b), where the additive formulation $\Snl$ shows a significantly sharper and higher peak compared to either damping only or the rigorously derived $\Snld,$ which again yields a lower and broader spectrum. Once the damping is comparable to the initial rate of nonlinear interactions, differences between the formulations are largely erased, as can be seen in Figure \ref{fig:spectra-equal-energy}(c). Using the Hellinger distance to quantify these spectral shape changes, we find $d_H$ between rigorous and additive simulations reduces from $0.20$ to $0.03$ as the damping rate is increased from $\mathcal{O}(\epsilon_0^4\omega_{p,0})$ to $\mathcal{O}(\epsilon_0^2\omega_{p,0})$.

When phase-mixing is introduced into the simulations (dashed blue and red), we see that it has a regularising effect irrespective of the choice of $\Snl$ or $\Snld$. Indeed, in all three panels of Figure \ref{fig:spectra-equal-energy}, the spectra with phase-mixing nearly coincide. The rate of phase mixing is fixed at $50T_{p,0}$, and so the weakest damping and longest simulations (with $t^*=585 T_{p,0}$) shown in panel (a) undergo more phase mixing events than those in panels (b) and (c) (with $t^*=99 T_{p,0}$ and $17 T_{p,0}$, respectively), where the more rapid damping of the spectrum leads to an earlier termination of our simulation. This explains why the phase-mixed simulations in panel (a) lead to significant broadening of the spectrum and down-shifting of the spectral peak compared with the other cases considered.

\begin{figure}
    \centering
    \includegraphics[width=\textwidth]{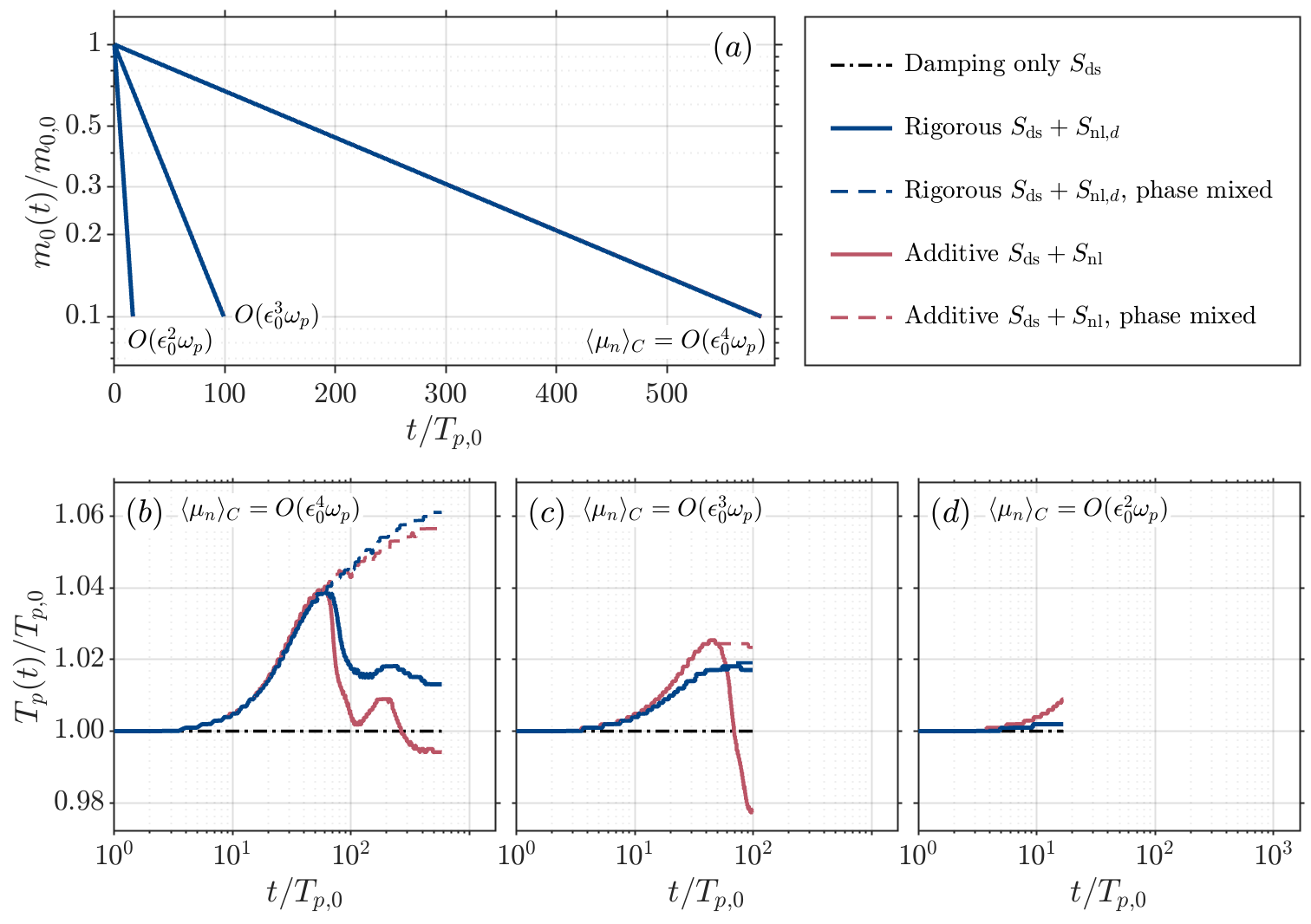}
    \caption{(a) Temporal evolution of the dimensionless energy, $ m_0 $, and (b--d) normalised peak period in panels for uniform damping in panels:
    (b) $ \langle\mu_n \rangle_C  = 1.6 \epsilon_0^4 \omega_{p,0}$; (c) $\langle\mu_n \rangle_C = 1.1 \epsilon_0^3\omega_{p,0}$; (d) $\langle\mu_n \rangle_C = 0.79 \epsilon_0 ^2\omega_{p,0}$. A vertical logarithmic axis is used in (a) and logarithmic horizontal axis in panels (b--d).}
    \label{fig:moment-and-peak-period}
\end{figure}

Despite the differences in spectral shape evident from the time-evolution, the total energy of the spectra $m_0$ varies very little among the formulations (with a maximum deviation of 0.48\%), as shown in Figure \ref{fig:moment-and-peak-period}(a) where the lines for the different formulations overlap for a prescribed damping rate. The spectral significant wave-height -- derived from $m_0$ -- is therefore also insensitive to the formulation used. Figure \ref{fig:spectra-equal-energy} makes clear that the peak period $T_p$ can shift significantly over the evolution, as is shown in more detail Figure \ref{fig:moment-and-peak-period}(b)-(d) for the three damping rates considered.

The normalised peak period $T_p/T_{p,0}$ shows a fast, initial increase in all cases when the nonlinear term is included, coincident with the fast, near-resonant-interaction time-scale $\mathcal{O}(\epsilon_0^{-2}{\omega_{p,0}}^{-1})$ of the GKE evolution. In the absence of phase-mixing (continuous red and blue), this increase in peak period is followed by a decrease which characterises the subsequent evolution, whose end state at $t^*$ is shown in Figure \ref{fig:spectra-equal-energy}. Recalling that the damping rate itself is incorporated into the rigorous $\Snld$ term, it is no surprise that the spectral peak of the additive formulation with $\Snl$ departs from the rigorous counterpart at ever earlier times as the dissipation rate is increased from $\mathcal{O}(\epsilon_0^4 \omega_{p,0})$ to $\mathcal{O}(\epsilon_0^2 \omega_{p,0}),$ as shown in panels (b)--(d) of Figure \ref{fig:moment-and-peak-period}. 

After the initial spectral downshift, the additive formulation $\Snl$ (dashed lines) is seen to lead to a fast up--shift which eventually overshoots the initial spectral peak for cases (b) and (c). By contrast, the rigorous formulation of $\Snld$ retains a downshifted spectral peak, with the peak adjustment appearing to flatten off in the absence of phase-mixing. When phase mixing is included, the fast, initial adjustment of the GKE is restarted every $50 T_{p,0},$ which induces the spectral down-shifting to continue throughout the simulation in case (b), while also leading to a smaller but noticeable increase in peak period for the stronger damping of case (c).

\subsection{Time Scales}
\label{ssec:Time scales}

The shifts in regimes evident in Figures \ref{fig:spectra-equal-energy} and \ref{fig:moment-and-peak-period} can be related to the interaction of different time-scales in the problem. While the dissipation rate -- and its associated time-scale -- is constant, the generalised kinetic equation embodies both a fast $\mathcal{O}(\epsilon_0^{-2} \omega_{p,0}^{-1})$ adjustment (associated with the memory integral in \eqref{eq:dGKE}) and a slower, $\mathcal{O}(\epsilon_0^{-4}\omega_{p,0}^{-1})$ kinetic time-scale. The action of dissipation means that nonlinearity steadily decreases from $\epsilon_0$, so that the time-scales associated with $\Snl$ change throughout the problem. This change is not monotonic due to the competing effect of nonlinear interactions.

Recalling the schematic form of our damped generalised kinetic equation $ C_n' = \Sds + \Snld$ (or the related equation with purely additive damping $C_n' = \Sds + \Snl$), we  define the instanteneous nonlinear and dissipative time scales, denoted $\tau_{nl}$, $\tau_{nl,d}$ and $\tau_{ds}$, respectively as
\begin{equation}
    \label{eq:time-scales}
\tau_{nl}(t)=\frac{\int C\,\mathrm{d}k}{\int |\Snl|\,\mathrm{d}k}, \qquad \tau_{nl,d}(t)=\frac{\int C\,\mathrm{d}k}{\int |\Snld|\,\mathrm{d}k}, \qquad \tau_{ds}=\frac{\int C\,\mathrm{d}k}{\int |\Sds| \,\mathrm{d}k},
\end{equation}
where the simple form of $\Sds=-2\mu C_n$  ensures that $\tau_{ds}$ is independent of time.

\begin{figure}
    \centering
    \includegraphics[width=\textwidth]{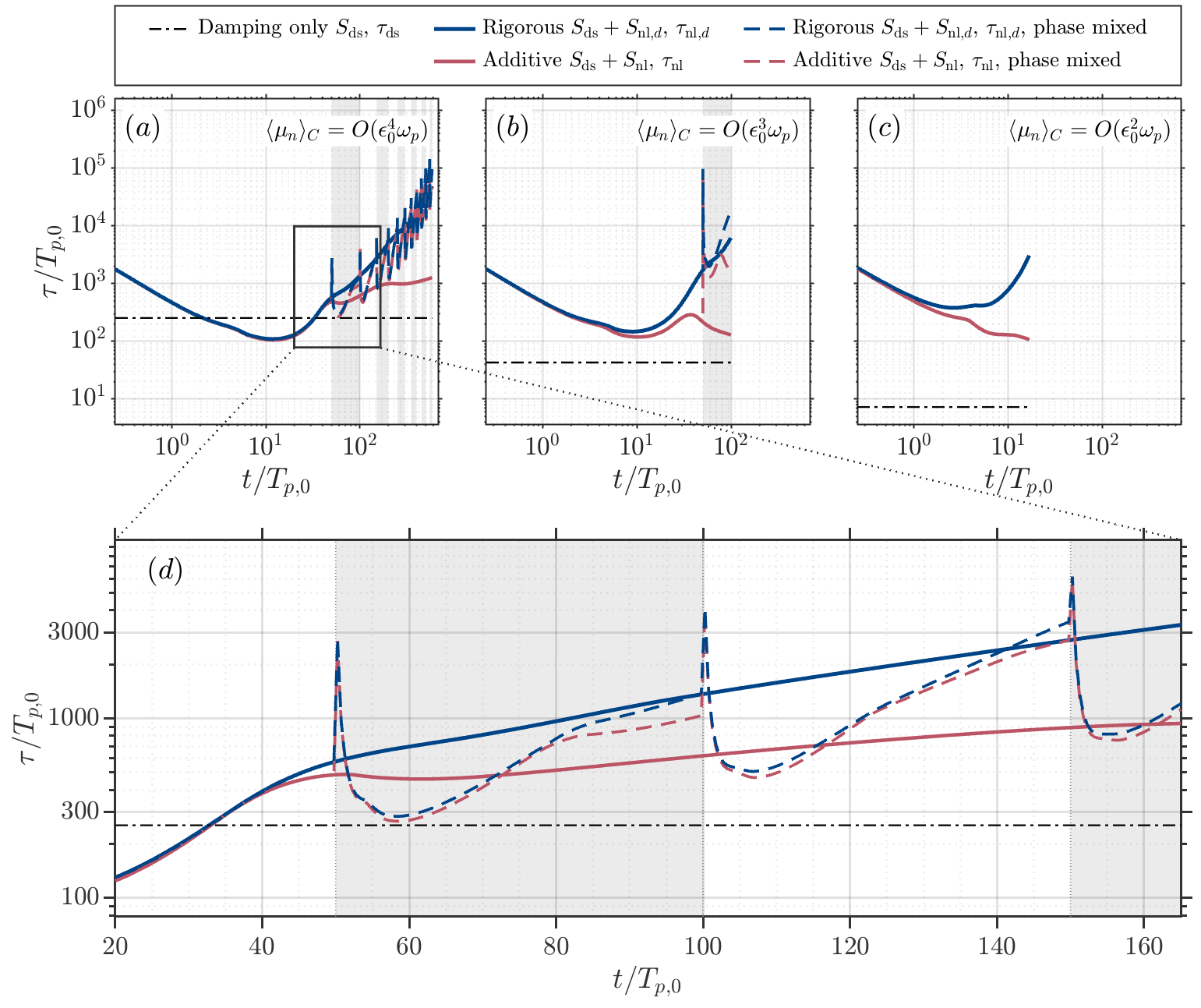}
    \caption{Temporal evolution of the damping and nonlinear time scales \eqref{eq:time-scales} for uniform damping: (a) $\langle\mu_n \rangle_C  = 1.6 \epsilon_0^4 \omega_{p,0}$; (b) $\langle\mu_n \rangle_C = 1.1 \epsilon_0^3\omega_{p,0}$; (c) $\langle\mu_n \rangle_C = 0.79 \epsilon_0 ^2\omega_{p,0}$ . Axes are logarithmic in (a)--(c). A zoom in on panel (a) is shown in panel (d) with linear time horizontal axis. The alternating grey-transparent background highlights the phase mixing interval $50T_{p,0}$ used for simulations. }
    \label{fig:time-scales-and-zoom-in}
\end{figure}

Figure \ref{fig:time-scales-and-zoom-in} shows the evolution of the local nonlinear and dissipative time scales for all the cases considered. Figure \ref{fig:time-scales-and-zoom-in}(a), with the smallest dissipation rate, shows that the nonlinear times scales quickly become smaller (faster) than the dissipative time scales, and remain so until approximately $33 T_{p,0}$, coinciding with the fast adjustment and downshift observed in Figure \ref{fig:moment-and-peak-period}(b). Soon after this adjustment, the different nonlinear time-scales diverge -- shown magnified in panel (d) -- with both $\tau_{nl}$ and $\tau_{nl,d}$ growing exponentially, albeit at different rates. This difference can be directly attributed to the memory term, which is independent of the damping for $\Snl$, but contains a dissipative correction in $\Snld$.  When phase-mixing is applied, each phase-mixing event re-starts an attendant fast adjustment in the time-scales $\tau_{nl}$ and $\tau_{nl,d}$, similar in form to the initial evolution, as can be seen in panel (d). Interestingly, the long-time behaviour of both phase-mixed time scales more closely resembles the rigorous $\tau_{nl,d}$ without phase mixing than it does the purely additive, non-phase mixed scale $\tau_{nl}.$

For stronger damping shown in Figure \ref{fig:time-scales-and-zoom-in}(b),(c) the dissipative time-scale $\tau_{ds}$ is faster than the nonlinear time-scales for all cases considered. While the rigorous $\tau_{nl,d}$ scale exhibits a minimum at each of the three damping rates -- predominantly driven by the nonlinear interaction and with location of the minimum only weakly influenced by the damping rate itself -- the additive $\tau_{nl}$ scale does not. This is another manifestation of the decoupling of nonlinearity and dissipation, whereby the memory integral governing $\tau_{nl,d}$ gradually loses information about past states due to the inclusion of the exponential damping term $\Delta \mu_{npqr}$ in \eqref{eq:dGKE}, in contrast to the conservative memory integral of \eqref{eq:additive-GKE} which gives rise to $\tau_{nl}$.

\subsection{Quantifying spectral changes with dissipation strength}
\label{ssec:varying-forcing}

We have observed in Figure \ref{fig:spectra-equal-energy} that varying dissipation rates  influence the extent of spectral shape changes in the additive and rigorous formulations, as does the inclusion of phase mixing. To extensively quantify these changes, we consider a range of dissipation rates between $\langle \mu_n \rangle_C =2.35 \times  10^{-5} \text{s}^{-1}$ and $1 \text{s}^{-1}$, corresponding to $2.1 \epsilon_0^5 \omega_{p,0}$ and $2.3 \epsilon_0^0 \omega_{p,0}$, and simulate the evolution of the initial spectrum given in Section~\ref{subsec:numerical-implementation} until time $ t=t^* $, see \eqref{eq:tStar-def}, with and without phase mixing every $50 T_{p,0}.$

\begin{figure}
    \centering
    \includegraphics[width=\textwidth]{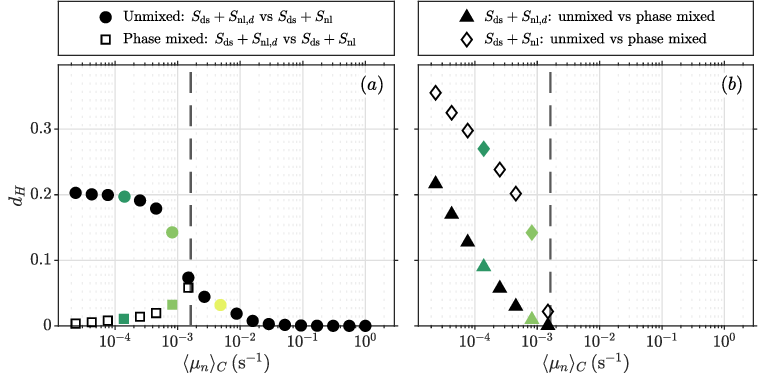}
    \caption{Hellinger distance $d_H$ at $t^*$ for varying (uniform) damping rates. (a) rigorous vs additive formulation with (squares) and without (circles) phase mixing. (b) phase-mixed vs unmixed, for the rigorous (triangles) and additive (diamonds) formulation.  The three damping rates considered in Figures \ref{fig:spectra-equal-energy}--\ref{fig:time-scales-and-zoom-in} are highlighted in shades of green. The grey, dashed line denotes $\mu_{n,\text{crit}}.$}
    \label{fig:Hellinger}
\end{figure}

We measure spectral shape changes by means of Hellinger distance \eqref{eq:Hellinger-def}, recalling that $d_H$ ranges from 0 for identical normalised spectra to 1 for non-overlapping normalised spectra. Figure \ref{fig:Hellinger}(a) compares the rigorous and additive formulations. In the absence of phase mixing (circular markers), smaller dissipation rates lead to greater dissimilarities in spectral shape between $\Snl$ and $\Snld.$ As dissipation increases, the spectra become increasingly similar because of the dominant effect of the linear damping term over the nonlinear term -- as seen also in Figure \ref{fig:spectra-equal-energy}(c) -- with the two formulations becoming indistinguishable for $\langle \mu_n \rangle_C \sim 0.1  \text{s}^{-1}$. This situation is reversed in simulations with phase mixing (square markers), although overall differences in spectral shape are smaller. Indeed, for weak dissipation, phase-mixing tends to diminish differences between the additive and rigorous formulations. Note that for dissipation rates larger than $ \mu_{n,\text{crit}} = \frac{1}{2}T_r^{-1}\ln(10)=1.62 \times 10^{-3} \text{s}^{-1} \approx2.1 \epsilon_0^3\omega_{p,0}$ (dashed line in Figure \ref{fig:Hellinger}), the decline in total energy is so rapid that no phase mixing can take place.   

When comparing the effects of phase mixing within each individual formulation, Figure \ref{fig:Hellinger}(b), we observe that phase-mixing leads to larger changes in spectral shape for the additive formulation $\Snl$ than the rigorous $\Snld$, with diamond markers consistently lying above triangle markers for all damping rates considered. This reflects the tendency observed in Figure (\ref{fig:spectra-equal-energy})(a), where we observed a lower, broader peak in simulations of $\Snld$ both with and without phase-mixing, contrasting with the sharper, higher peak in $\Snl$ simulations without phase mixing. As the damping rate increases, Figure \ref{fig:Hellinger}(b) demonstrates that differences in spectral shape induced by phase mixing rapidly diminish, irrespective of the formulation used. For damping in excess of $ \mu_{n,\text{crit}}$ (which includes the case $\langle \mu_n \rangle_C = 0.79 \epsilon_0 ^2\omega_{p,0}$) our simulation terminates before phase mixing can take place, so no comparison is possible.

\section{Results for frequency-dependent damping}
\label{sec:frequency-dependent-damping}

In Section \ref{sec:uniform-damping-results} we investigated an idealised type of frequency-independent dissipation, namely constant dissipation over the entire spectrum. Physically, realistic modelling of dissipation mechanisms in a spectral model generally requires the dissipation to vary with wave frequency \citep{KomenEtAl}. Depending on the mechanism of dissipation, this frequency dependence may take different forms. For example, wave sea-ice interactions are often modelled as a dissipation $\mu_n \propto \omega_n^N$ \citep{meylan2018jgr,Ransome2026}. In this section, we will employ a frequency--dependent power law damping of the form
\begin{equation} \label{eq:omega2-damping}
\mu_n=\beta\omega_n^{2}.
\end{equation}
Values of the coefficient $\beta$ are chosen so that the averaged damping rates defined by \eqref{eq:forcing-average} agree with those presented in Section~\ref{subsec:numerical-implementation}: $\langle \mu_n \rangle _C  = 1.6 \epsilon_0^4 \omega_{p,0} , 1.1 \epsilon_0^3\omega_{p,0} , 0.79 \epsilon_0 ^2\omega_{p,0}$. The numerical configuration is also the same as outlined in Section~\ref{subsec:numerical-implementation}, with final simulation times $t^* = 636 T_{p,0}, 108T_{p,0} $ and $18.2T_{p,0}$ respectively corresponding to a diminution of total spectral energy to $10\% $ of its initial value, {see \eqref{eq:tStar-def}}. 

\begin{figure}
    \centering
    \includegraphics[width=\textwidth]{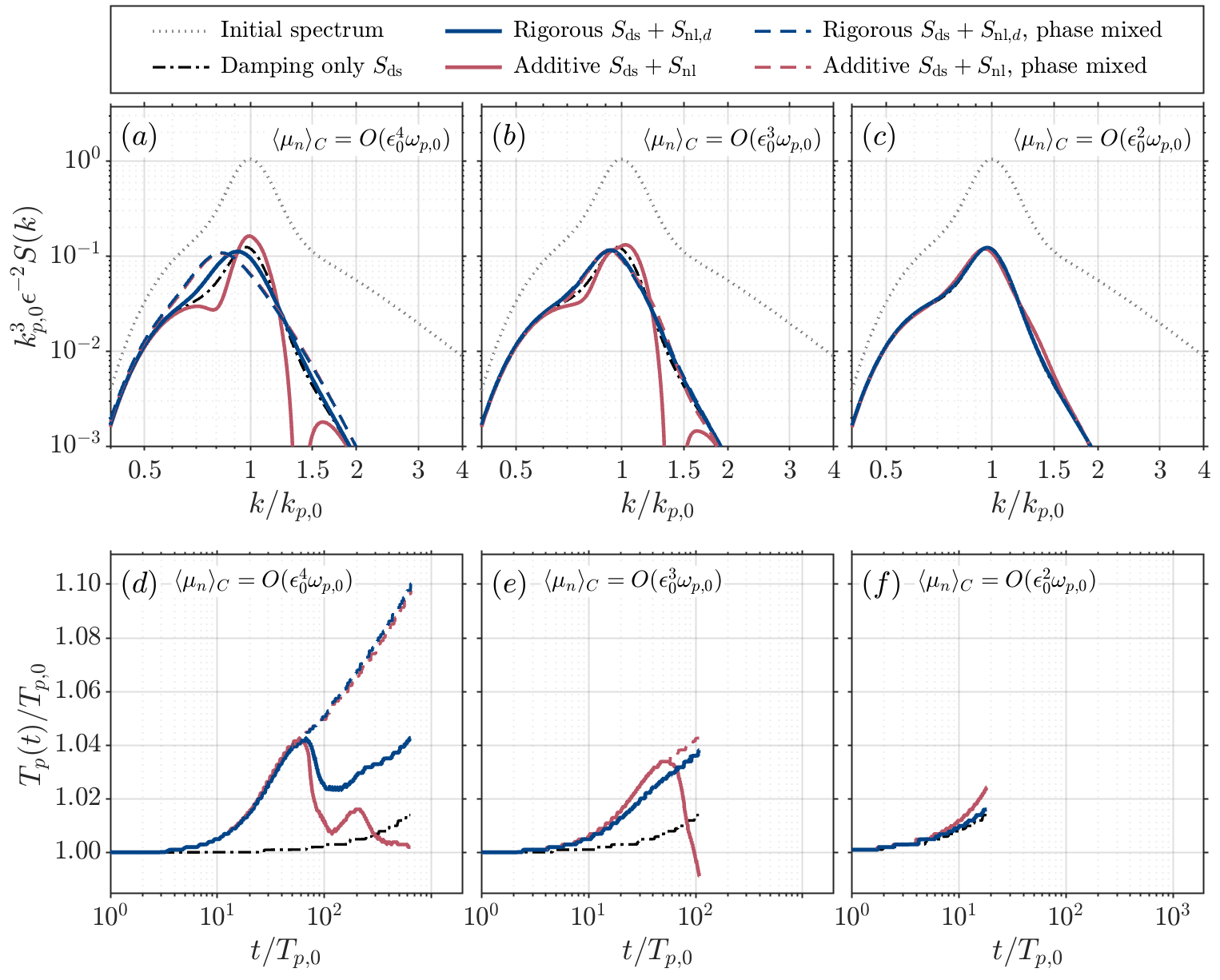}
    \caption{Dimensionless wavenumber spectra at $t^*$ (a--c) and normalised peak period (d--f) for frequency-dependent damping $\mu_n 
    \propto \omega_n^2 $: (a; d)  $\langle\mu_n \rangle_C  = 1.6 \epsilon_0^4 \omega_{p,0}$, (b;e) $\langle\mu_n \rangle_C = 1.1 \epsilon_0^3\omega_{p,0}$ , (c; f) $\langle\mu_n \rangle_C = 0.79 \epsilon_0 ^2\omega_{p,0}$. }
    \label{fig:spectra-equal-energy-omega2}
\end{figure}

Unlike in Section \ref{sec:uniform-damping-results}, frequency--dependent damping itself causes changes in spectral shape as shown in the dash-dotted black curves in Figure \ref{fig:spectra-equal-energy-omega2} (a)--(c), a fact which is also reflected in the peak downshift visible for the damping only cases shown in panels (d)--(f).

For the two weaker dissipation cases shown in panels (a) \& (b), the purely additive formulation of the nonlinear source term $\Snl$ exhibits markedly unphysical behaviour, with the spectra showing gaps in wavenumber space, in addition to presenting the highly peaked form already observed for uniform damping in Figure \ref{fig:spectra-equal-energy}. By contrast, the rigorous formulation $\Snld$ shows a broad shape with downshifted peak which is intermediate between the phase-mixed simulations and the damping-only case. 

In all cases the high-frequency tail of the spectrum is preferentially damped, so that spectral energy is essentially negligible beyond $2 k_{p,0}$ (cf.\ Figure \ref{fig:spectra-equal-energy}). The failure of the additive formulation to properly account for damping leads to deviations between this and other formulations in the large--wavenumber range, even for the very high dissipation rates shown in Figure \ref{fig:spectra-equal-energy-omega2}(c). Nevertheless, all formulations, with or without phase-mixing, approach a similar evolution for sufficiently high dissipation, mirroring what we observed for uniform damping in Figure \ref{fig:Hellinger}.

When phase-mixing is incorporated, the difference between $\Snl$ and $\Snld$ is nearly negligible in all cases. Resetting the phases every $50 T_{p,0}$ leads to a monotonic downshift of the spectral peak over the entire evolution, shown in Figure \ref{fig:spectra-equal-energy-omega2}(d)--(f). Without phase mixing, the peak downshift arising from fast, near-resonant interactions is arrested as slower, resonant dynamics take hold and vie with energy dissipation to reshape the spectrum.

\section{Discussion and conclusions}
\label{sec:Discussion and conclusions}

Dissipative processes in surface wave propagation such as whitecapping are strongly localised and highly nonlinear. However, because their overall energy transfer per period is small for open-ocean propagation, when these processes are incorporated into spectral wave models they appear as quasi-linear terms in the spectrum.  These linear damping terms are usually added \textit{a posteriori}, and this addition is subsequently justified by appealing to a hierarchy of time-scales.
This body of ideas has recently been critically re-examined by \citet{maestrini2025wave}, who derived a kinetic equation directly from the damped, deterministic  Zakharov equation (see also \cite{lin2026wave}). If the nonlinearity and dissipation are assumed to act on the same dynamical time-scale, they found that this requires changes to the collision integral. 

We set out to probe the separation of damping and nonlinear interaction via a rigorous derivation of a generalised kinetic equation. In contrast to the kinetic equation considered by \citet{maestrini2025wave}, which describes spectral evolution due to exact resonant interactions on the long time scale $\mathcal{O}(T/\epsilon_0^{4})$, the generalised kinetic equation also incorporates near-resonant interactions and so exhibits an additional fast evolution $\mathcal{O}(T/\epsilon_0^{2}).$ This allows for a rich interplay between nonlinear and dissipative scales, whose behaviour we capture by numerical simulations of the respective equations.

The generalised kinetic equation we employ builds on work by \citet{Janssen2003}, \citet{Annenkov2006} and \citet{Gramstad2013}, and may be viewed as a generalisation of the classical kinetic equation of \citet{Hasselmann1962a}. Indeed, if we assume that the nonlinear term $f_{npqr}$ evolves on a sufficiently slow time-scale, it may be extracted from the time-integral in \eqref{eq:dGKE}. The subsequent integral can be evaluated to yield either Janssen's modified kinetic equation \citep{Janssen2003} if the wave field is assumed uncorrelated at a finite initial time, or Hasselmann's equation if the spectral components are assumed uncorrelated as $t \rightarrow -\infty.$ In the latter case, the integral collapses to a Dirac delta-distribution $\delta(\omega_n + \omega_p - \omega_q - \omega_r),$ enforcing the exact resonance condition. Analogous arguments allow for the reduction of our damped generalised kinetic equation \eqref{eq:dGKE} to that obtained by \citet{maestrini2025wave}.

When damping is incorporated \textit{ab initio} in the deterministic primitive equation, it appears as a linear term in the wave spectrum in the generalised kinetic equation, but also modifies the memory integral through inclusion of a new, exponential-decay term. This has the effect that memory of past states slowly fades away under influence of damping, and explains the regularising effect that our new nonlinear term exhibits in comparison to the inclusion of only a simple, additive damping. At the same time, as nonlinearity is proportional to wave steepness, the interplay between dissipation and nonlinear interaction is complex and defies a simple classification based only on initial conditions. If dissipation is not too strong, the fast, nonlinear dynamics of the generalised kinetic equation may constitute the dominant time-scale during the initial evolution of the spectrum. In the absence of energy input, at sufficiently long times the dynamics will inevitably transition to a dissipation-dominated regime where nonlinearity plays a subordinated role. 

In the propagation of ocean waves, it is often assumed that wave breaking acts to de-correlate wave phases \citep{babanin2010numerical}. As the generalised kinetic equation allows for the evolution of fourth-order cumulants, which represent departures of the wave field from Gaussianity, it is common in numerical simulations to incorporate phase-mixing to effect such a de-correlation \citep{Gramstad2013,Gramstad2016,Liu_Gramstad_Babanin_2021}. When such phase-mixing is included in our simulations, it is seen to restart the initial fast, nonlinear evolution of the GKE, and to amplify spectral downshifting and broadening versus our rigorously derived equation. Phase mixing has the additional benefit of ameliorating numerical instabilities in discretised versions of the generalised kinetic equation, which have been reported in a number of recent works \cite[p.\ 911]{Liu_Gramstad_Babanin_2021} (see also \cite[Fig.\ 14]{Liu_Gramstad_Babanin_2021}, \cite[Fig.\ 16]{Gramstad2013}, and \cite{Andrade2019}). 

One major advantage of our damped generalised kinetic equation is that it does not exhibit such instabilities. In simulations up to 2000 $T_{p,0}$, our newly derived equation showed robust spectral evolution. In contrast, inclusion of a purely additive damping term in the generalised kinetic equation can be seen in Figures \ref{fig:spectra-equal-energy} and \ref{fig:spectra-equal-energy-omega2} to produce multiply peaked spectra and unphysical evolution unless phase-mixing is included. The favourable stability properties of our formulation may be attributed to the transience in the memory term introduced by the rigorous inclusion of dissipation.

Although our new, damped GKE is applicable to directional spectra, we have restricted our simulations to the computationally simpler setting of unidirectional waves. In addition, phase-mixing has been incorporated on an \textit{ad hoc} basis to allow for comparison with previous work \citep{Gramstad2013}, and damping has been restricted to two idealised cases. Further work is therefore necessary to assess the inclusion of our new formulation in wave models, including studying the effects of wind-forcing. For some settings, such as waves propagating into sea-ice \citep{Alberello2022NatComms,Stuhlmeier2024modulational}, damping due to wave-ice interactions should be modelled using a spatial rather than temporal evolution, with attendant modifications. 

Our new, consistently derived generalised kinetic equation does not require phase mixing for numerical stability. This isolates the dual purposes of phase mixing, allowing it to be incorporated when physically meaningful -- such as in certain frequency domains -- while omitting it when it is not, without negatively impacting the long--time fidelity of simulations. This circumstance also opens up the possibility of exploring couplings between dissipative mechanisms and mixing rates, and thereby introducing more realistic physics of ocean waves. While this work has focused on damping only, the formulation presented applies also to forcing. The results presented herein therefore show promise for future implementation of generalised kinetic equations in wave models and coupled Earth system models.

\section*{Acknowledgements}

R.R. is supported by the Natural Environment Research Council and the ARIES Doctoral Training Partnership [grant number NE/S007334/1].
R.R. and A.A. acknowledge funding from Royal Society (IEC/R3/243016).
D.P. is supported by EPSRC Grant No. EP/Y021118/1 and by the ExtreMe Matter Institute EMMI at the GSI Helmholtzzentrum fuer Schwerionenphysik, Darmstadt, Germany.
This work was stimulated by discussions during the ‘Maths of Sea Ice’ meeting, made possible by the Isaac Newton Institute for Mathematical Sciences.

\end{document}